\documentclass[twocolumn,showpacs,preprintnumbers,amsmath,amssymb]{revtex4}
\usepackage{graphicx}
\usepackage{dcolumn}
\usepackage{bm}

\DeclareGraphicsExtensions{eps}

\begin{document}
\title{Incommensurate spin density wave in Co-doped BaFe$_2$As$_2$}
\author{P. Bonville, F. Rullier-Albenque, D. Colson, A. Forget}

\affiliation{CEA, Centre de Saclay, DSM/IRAMIS/Service de Physique de l'Etat Condens\'e\\
91191 Gif-sur-Yvette, France}

\begin{abstract}
$^{57}$Fe M\"ossbauer spectroscopy measurements are presented in the
underdoped Ba(Fe$_{1-x}$Co$_x$)$_2$As$_2$ series for $x$=0.014 ($T_c <
1.4$\,K) and
$x$=0.03 and 0.045 ($T_c \simeq$ 2 and 12\,K respectively). The spectral
shapes in the
so-called spin-density wave (SDW) phase are interpreted in terms
of incommensurate modulation of the magnetic structure, and allow the shape
of the modulation to be determined. In undoped BaFe$_2$As$_2$, the magnetic
structure is commensurate, and we find that incommensurability is present
at the lowest doping level ($x$=0.014). As Co doping increases, the low
temperature modulation progressively loses its ``squaredness'' and tends
to a sine-wave. The same trend occurs for a given doping level, as temperature
increases. We find that a magnetic hyperfine component persists far above the
SDW transition, its intensity being progressively tranferred to a paramagnetic
component on heating.
\end{abstract}

\maketitle

\section{Introduction}

Investigation of the magnetic properties of the newly discovered layered
Fe-based pnictide supercondutors \cite{kami} with a microscopic probe such
as M\"ossbauer spectroscopy is of interest for (at least) two reasons:
i) to determine the characteristics of the expected spin-density wave (SDW)
order and ii) to give experimental evidence concerning the problem of the
interaction and/or local coexistence between superconductivity and magnetic
ordering of the Fe moments. In the so-called 122 family with Ba, the parent
(non superconducting) compound is BaFe$_2$As$_2$ and doping with Co
substituted
for Fe has been shown to induce superconductivity at low doping levels
\cite{sefat}. In the Ba(Fe$_{1-x}$Co$_x$)$_2$As$_2$ series, the phase diagram
as a function of Co concentration $x$ is ``bell-shaped'', with $T_c(x)$
presenting a maximum of 26\,K for $x$=0.07 \cite{chu,flo}. Below this optimal
concentration, the material shows
apparently the persistence of an antiferromagnetic (AF) structure in the
superconducting phase, the AF or spin density wave (SDW) transition
temperature decreasing rapidly as $x$ increases. Neutron diffraction in the
undoped compound BaFe$_2$As$_2$ \cite{huang} has determined the AF magnetic
structure below $T_N \simeq$143\,K to be {\it a priori} commensurate.
$^{57}$Fe M\"ossbauer spectroscopy \cite{rotter0,rotter} and $^{75}$As NMR
\cite{kita} in BaFe$_2$As$_2$ have confirmed the commensurability of the
magnetic structure.
\begin{figure}
\includegraphics[width=11cm]{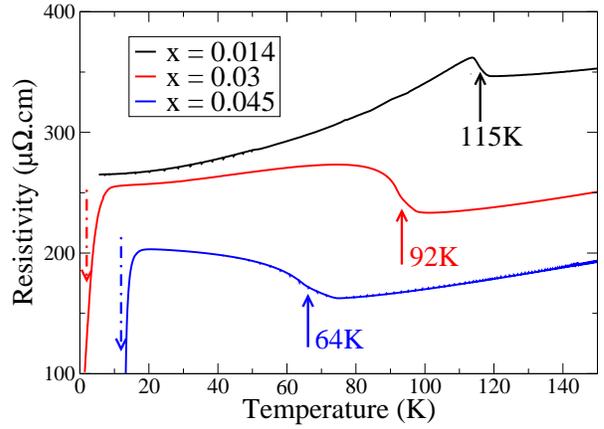}
\caption{[color online] Resistivity {\it vs.} temperature for BaFe$_2$As$_2$
with Co dopings $x$=0.014, 0.03 and
0.045 from Ref.\cite{flo}. The data for $x$=0.014 have been shifted by
100\,$\mu\Omega$.cm for clarity. The full arrows mark the inflexion point of
$\rho(T)$, attributed to the onset of magnetic ordering ($T_{\rm SDW}$), and
the dash-dotted arrows the superconducting transition temperature.}
\label{rhoT}
\end{figure}

In this work, we focus on the first problem evoked above: we performed a
$^{57}$Fe M\"ossbauer study of Ba(Fe$_{1-x}$Co$_x$)$_2$As$_2$ in the
underdoped region in order to investigate the SDW state. In our samples with
Co doping $x$=0.014, 0.03 and 0.045, the resistivity data show an inflexion
point \cite{flo}
respectively around 115\,K, 92\,K and 64\,K (see Fig.\ref{rhoT}). According to
neutron diffraction data in materials with similar
Co dopings \cite{pratt,christianson,lester}, this temperature corresponds
to the SDW-paramagnetic transition. We find that the spectra in the magnetic
phase have an unusual shape, which we interpret as due to an incommensurate
modulation of the AF structure, i.e. to an incommensurate (IC) spin density
wave.
We also find that the magnetic hyperfine subspectrum persists far above
$T_{\rm SDW}$, i.e. that AF correlations are present up to $T_{\rm AF} \sim
1.5 \ T_{\rm SDW}$.

\section{Experimental and sample characterisation}

The samples are single crystals prepared by the self-flux method, as described
in Ref.\cite{flo}. It was checked that Co doping is uniform and that
superconductivity occurs in the bulk \cite{lapl,lapl2}. For the M\"ossbauer
measurements, the crystals were
finely ground. Powder samples made by solid state reaction were also prepared,
with $x$=0 and 0.03, for comparison purposes. The spectra were recorded using
a commercial
$^{57}$Co$^*$:Rh $\gamma$-ray source mounted on an electromagnetic drive with
linear velocity signal.
Room temperature spectra ensured that no trace of FeAs (with a quadrupole
splitting $\Delta E_Q$=0.69\,mm/s) was present.

In the high temperature tetragonal phase (space group $I4/mmm$), the $4d$ Fe
site has fourfold symmetry ($\bar 4$m2). At low temperature, slightly above
$T_{\rm SDW}$, a
small orthorhombic distortion takes place (space group $Fmmm$), and the
local symmetry at the $8f$ Fe site is lowered (222).

\section{The $^{57}$Fe M\"ossbauer spectra at 4.2\,K}

\begin{figure}
\includegraphics[width=9cm]{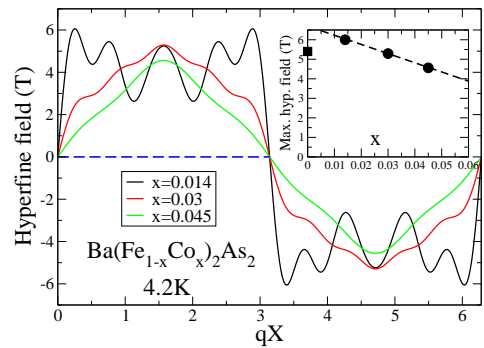}
\label{spe4k2Co}
\caption{[color online] Modulation of the hyperfine field along the direction
of {\bf q} at 4.2\,K in
Ba(Fe$_{1-x}$Co$_x$)$_2$As$_2$ for $x$=0.014, 0.03 and
0.045, derived from the M\"ossbauer spectra. Insert: variation with the Co
concentration $x$ of the maximum hyperfine field value (circles); for $x$=0,
hyperfine field in BaFe$_2$As$_2$ (square). The dashed line is a guide for
the eye.}
\label{mod4k2}
\end{figure}

The spectra at 4.2\,K in the three investigated compounds and in undoped
BaFe$_2$As$_2$, are represented in
Fig.~\ref{spe4k2Co}. In BaFe$_2$As$_2$, one observes a single magnetic
hyperfine pattern, in agreement with Ref.\cite{rotter}, with a hyperfine field
$H_{\rm hf} = 5.4(1)$\,T. By contrast, the spectral shapes in the Co-doped
compounds are rather far from this standard 6-line pattern.
Saturation effects can be discarded since the resonant
absorption is modest; for $x$=0.03, where it amounts to 7\%, we checked
that an absorber with 2\% absorption yields exactly the same spectrum.
Texture effects are also negligible since a powder absorber (for $x$=0.03)
yields spectra quasi-identical with those of the single crystal sample. The
high statistics reached in these experiments allows us to distinguish fine
details
of the spectral shape, which is a prerequisite for the fits to be described
below.

\begin{table}
\caption{In Ba(Fe$_{1-x}$Co$_x$)$_2$As$_2$: Fourier coefficients $h_{2k+1}$
and maximum value $h_m$ (in T) of the hyperfine field modulation at 4.2\,K .}
\label{coeff}
\begin{center}
\begin{tabular}{||c|c|c|c|c|c|c|c||} \hline
$x$   & $h_1$ & $h_3$ & $h_5$ & $h_7$ & $h_9$ & $h_{11}$ & $h_m$ \\ \hline
0.014 & 5.45 & 2.41 & 1.75 & 0.24 & 1.24 & 0.55 & 6.06\\ \hline
0.03  & 5.24 & 0.44 & 0.45 & 0.21 & 0.25 & -- & 5.29 \\ \hline
0.045 & 4.06 & -0.26 & 0.23 & -- & -- & -- & 4.55 \\ \hline
\end{tabular}
\end{center}
\end{table}
\begin{figure}
\includegraphics[width=6cm]{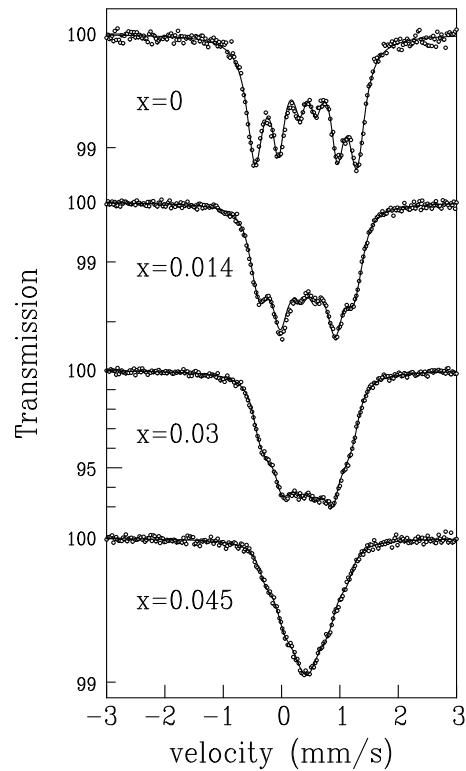}
\caption{M\"ossbauer absorption spectra on $^{57}$Fe at 4.2\,K
in Ba(Fe$_{1-x}$Co$_x$)$_2$As$_2$ for $x$=0, 0.014, 0.03 and 0.045. Except for
$x$=0, the solid
lines, which are masked by the data due to the goodness of the fits,
are simulations with an incommensurate modulation of hyperfine fields (see
text).}
\end{figure}
We interpret these unusual spectral shapes as due to a distribution of
hyperfine fields arising from an IC modulated magnetic structure.
Indeed, in this case, the $^{57}$Fe nuclei are submitted to hyperfine field
values ranging from zero to a maximum value, and although the M\"ossbauer
data cannot give access to the propagation vector {\bf q}, the shape of the
distribution reflects the particular shape of the modulation. In order to
describe the influence of the moment modulation on the spectral shape, we
assume a collinear magnetic structure, with no ferromagnetic component,
like in undoped BaFe$_2$As$_2$ where the Fe moments lie along the orthorhombic
{\bf a} axis \cite{huang}. We also assume the hyperfine field is collinear
with the Fe moment (see discussion below).
Then, along the direction OX parallel to the propagation vector {\bf q}, the
modulation of the hyperfine field is defined in terms of a Fourier series as:
\begin{equation}
H_{hf}(qX) = \sum_{k=0,n} h_{2k+1}\ \sin[(2k+1)qX],
\label{eqmod}
\end{equation}
where the $h_i$ are the odd Fourier coefficients of the modulation. The
spectra are then fitted to a superposition of individual hyperfine
6-line patterns with discrete $H_{hf}$ values according to Eq.(\ref{eqmod})
with a linear mesh along a period (0$\le qX\le 2\pi$). The number of Fourier
components in the fit is increased until a good reproduction of the lineshape
is obtained. At 4.2\,K, the very good fits shown in Fig.\ref{spe4k2Co} were
performed with 6, 5 and 3 Fourier components for $x$=0.014, 0.03 and 0.045
respectively. The resulting
hyperfine field modulations are shown in Fig.\ref{mod4k2}, and the values
of the Fourier coefficents are given in Table \ref{coeff}.
The isomer shift at 4.2\,K with respect to $\alpha$-Fe is 0.54(1)\,mm/s and
a very small quadrupolar interaction with $\Delta E_Q \simeq -$0.04(2)\,mm/s
is needed to reproduce the slight asymmetry of the spectra.

The 4.2\,K modulation is seen to depart more and more from ``squaredness''
(which occurs for $x$=0) as $x$ increases; for $x$=0.045, the modulation is
close to a pure
sine-wave. This is reflected in the number of Fourier coefficients needed
to reproduce the lineshape (see Table \ref{coeff}), which decreases as $x$
increases. The maximum value of the hyperfine field at 4.2\,K decreases
as $x$ increases (see insert in Fig.\ref{mod4k2}). For $x$=0.045, it is
75\% of the $x$=0.014 value.

\section{Thermal variation of the M\"ossbauer spectra}

\begin{figure}
\includegraphics[width=9cm]{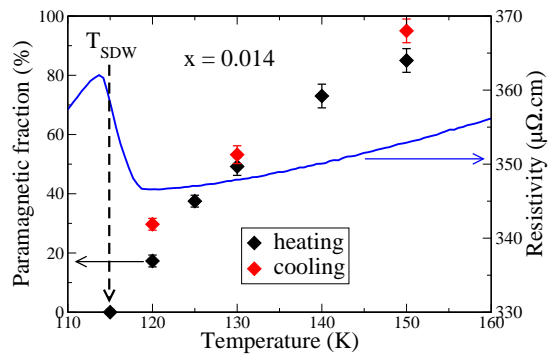}
\label{spe1.4CoT}
\caption{[color online] In BaFe$_2$As$_2$ with $x$=0.014 Co doping, thermal
variation of the
fraction of the single line component in the M\"ossbauer spectra, for
increasing and decreasing temperature runs (left scale), and of the
resistivity (right scale).}
\label{fracT}
\end{figure}

The thermal variation of the spectra for $x$=0.014, for which $T_{\rm SDW}
\simeq$115\,K (Fig.\ref{rhoT}), is shown in Fig.\ref{spe1.4CoT}. No important
change
occurs up to about 80K, then the modulation changes shape, tending towards
a pure sine-wave as temperature is further increased. A striking
feature is observed when comparing Fig.\ref{spe4k2Co} and Fig.\ref{spe1.4CoT}:
increasing the Co doping has the same effect on the modulation at 4.2\,K as
increasing the temperature for the $x$=0.014 sample.
The maximum value of the hyperfine field shows little thermal
variation, as well as the main Fourier coefficient $h_1$. Above
115\,K, a single
line, characteristic of the paramagnetic phase, is present in the spectra and
it grows on heating at the expense of the magnetic subspectrum (see red line
in the 130\,K spectrum in Fig.\ref{spe1.4CoT}). The latter actually
persists in a rather large temperature range above $T_{\rm SDW}$, up to
150\,K at least.
\begin{figure}
\includegraphics[width=6cm]{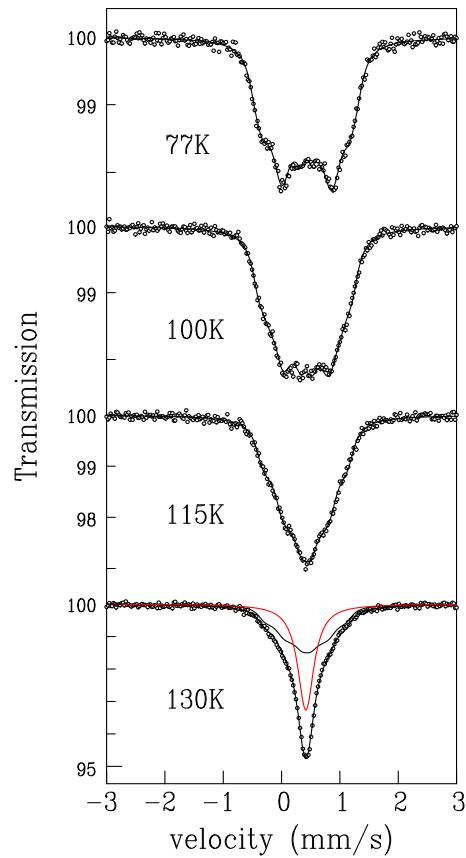}
\caption{[color on line] M\"ossbauer absorption spectra on $^{57}$Fe at
selected temperatures
in BaFe$_2$As$_2$ with $x$=0.014 Co doping. The lines are fits to an
incommensurate modulation of hyperfine fields; the red
subspectrum at 130\,K is a single Lorentzian-shaped line.}
\end{figure}
In this temperature range, very good fits of the spectra were obtained with
two components: a single Lorentzian-shaped line and a magnetic subspectrum
with a modulation identical to that of the 115\,K spectrum. For this latter
component, a common scaling factor of the Fourier coefficients was
introduced in order to allow for a decrease of the modulation amplitude on
heating; it remains above 0.9 up to the
highest temperature. The thermal variation of the fraction of the single line
derived from these fits is shown in Fig.\ref{fracT}: a small hysteresis is
observed and a unique line is recovered only above 150-160\,K.

\begin{figure}
\includegraphics[width=7cm]{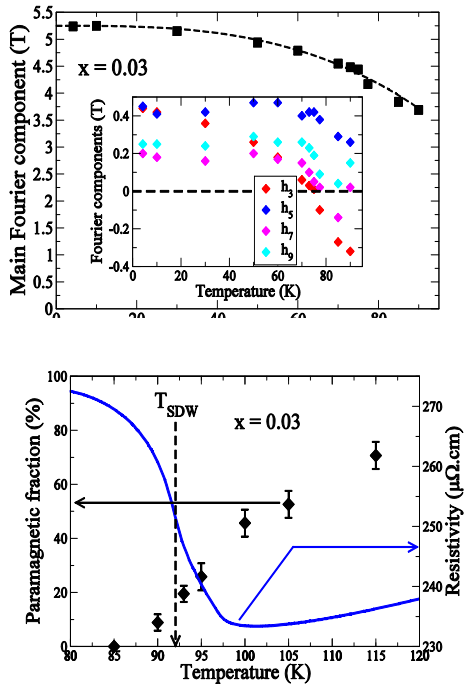}
\caption{[color online] In BaFe$_2$As$_2$ with $x$=0.03 Co doping: {\bf Top}
thermal variation of the main Fourier component $h_1$ of the hyperfine field
modulation (the line is a guide for the eye) and of the 4
next components $h_3 - h_9$ (insert) below $T_{\rm SDW}$ = 92\,K;
{\bf Bottom} thermal
variation (on heating) of the fraction of the single line (left scale) and
of the resistivity (right scale).}
\label{pourc3}
\end{figure}

For the sample with $x$=0.03, the thermal variation of the 5 Fourier
components was studied in more detail (see Fig.\ref{pourc3} top) and a similar
trend is observed. The main component $h_1$ is an order
of magnitude larger than the higher harmonics, and its variation with
temperature is weak: it falls by about 20\% between 4.2 and 90\,K, which is
close to $T_{\rm SDW}$ (see Fig.\ref{rhoT}).
The higher harmonics are constant up to 60\,K, then they drop and $h_3$
becomes negative and almost opposite to $h_5$, the other two harmonics being
smaller. In terms of the shape of the modulation, it does not change up to
about 60\,K, and on further heating it evolves smoothly towards an almost
pure sine-wave at 85\,K. The spectrum at 85\,K for $x$=0.03 is very
similar to the 4.2\,K spectrum for $x$=0.045, illustrating the above mentioned
trend that increasing $x$ at 4.2\,K has the same effect on the shape of the
modulation as increasing $T$
for a given $x$. Above 90\,K, i.e. above $T_{\rm SDW}$, a coexistence
region is present, like in the sample with $x$=0.014. The thermal variations
of the fraction of the single line and of the resistivity are represented in
Fig.\ref{pourc3} bottom. Above 115\,K, the two components of the M\"ossbauer
spectra cannot be distinguished, but the coexistence region probably extends
up to 130-140\,K.

For $x$=0.045, the modulation remains unchanged up to about 60\,K, close to
$T_{\rm SDW}$, then a single line grows in the spectrum on further heating.
However, due to the lack of resolution of the spectra, no quantitative
assessment of the fraction of single line can be done.
For this doping level,
the spectrum at 4.2\,K belongs to the superconducting phase, but we postpone
a discussion of the mixed phase (superconducting/SDW)
to a future publication.

According to the neutron diffraction data \cite{pratt,christianson,lester},
the transition to a long range ordered SDW state occurs at $T_{\rm SDW}$.
Actually, this transition is not ``seen'' in the spectra of
our underdoped samples; rather, {\it above $T_{\rm SDW}$}, a paramagnetic
component (single line) appears and coexists with the magnetic hyperfine
component. The latter progressively dwindles as temperature increases,
its (maximum) hyperfine field decreases slowly on heating, and so does the Fe
moment (see discussion below). We also observe a small hysteresis (for
$x$=0.014). This points to
the persistence of AF correlations, probably short range and dynamic, far
above $T_{\rm SDW}$. We estimate that a ``fully paramagnetic''
phase is recovered around $T_{\rm AF} \sim 1.5\  T_{\rm SDW}$.
This behaviour could be linked with the strong bidimensional character of the
magnetic interactions in the Fe layers \cite{wang,zhang}.

As to the structural transition which, in the Co doped compounds, occurs at
a temperature higher than $T_{\rm SDW}$ \cite{pratt,lester}, nothing can be
said from the M\"ossbauer data due to the very
weak quadrupolar interaction in these compounds.

\section{Discussion}
\subsection{Disorder induced by Co substitution}
Since these materials are (weakly) substituted compounds, one may wonder
whether the observed lineshapes, where a distribution of hyperfine
fields is clearly present, could be accounted for by some disorder induced
spread in the hyperfine field values.
\begin{figure}
\includegraphics[width=\linewidth]{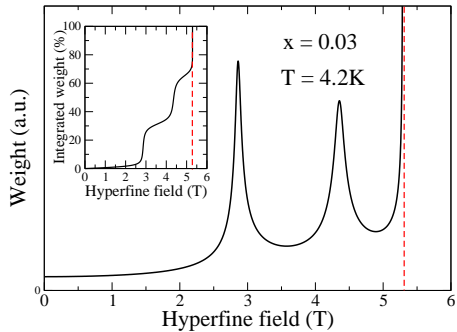}
\caption{[color online] Hyperfine field distribution associated with the
hyperfine field
modulation at 4.2\,K for the $x$=0.03 sample (see Table \ref{coeff}). The
positions of the peaks near 2.8 and 4.3\,T correspond to the quasi-plateaus
of the modulation, and the (unphysical) divergence to the maximum hyperfine
field where $dH_{hf}/d(qX)$=0 (see Fig.\ref{mod4k2}). Insert: integrated
weight.}
\label{disthf3}
\end{figure}
In the layered structure of the pnictides, a Fe atom has 4 in-plane nearest
neighbors, and for Co dopings $x$=0.03 and 0.045, it results that resp. 12\%
and 18\% Fe atoms have one Co atom as nearest neighbour. Assuming that the
hyperfine field is different for a Fe atom with 4 Fe and with 1 Co - 3 Fe as
nearest neighbors, then the corresponding M\"ossbauer spectrum would show a
two peak hyperfine field distribution with, for instance for $x$=0.03,
relative weights of 88 and 12\%. The distribution associated with the 4.2\,K
modulation for $x$=0.03 is shown in Fig.\ref{disthf3}: it presents
3 peaks, each with 25-30\% relative intensity, which therefore cannot
be due to the above hypothesis of different environments of a Fe atom.
The hyperfine fields of Fe atoms with 4 Fe or 1 Co - 3 Fe are probably very
close due to the delocalised nature of the 3$d$ electrons.

\subsection{Other evidences for incommensurability}
An incommensurability of the AF magnetic order in Co-underdoped BaFe$_2$As$_2$
has been inferred by NMR measurements \cite{ning,lapl} from the anisotropy of
the line broadening. The IC wave-vector was estimated in Ref.\cite{lapl} to
depart only slightly (by a few
percent) from the commensurate value {\bf q}=(101) in the orthorhombic cell.
$\mu$SR data in a sample with $x$=0.04 \cite{bernh} can also be
interpreted in terms of an IC magnetic structure.
The neutron diffraction measurements \cite{pratt,christianson,lester} did not
report any incommensurability, but the resolution in q-space is
probably not sufficient for this purpose; we note that the (a,b) orthorhombic
splitting could not be resolved in these experiments as well. In LaOFeAs, a
calculation of the Lindhardt response
function $\chi_0(q)$ was performed for the undoped system and for the system
with $x$=0.1 F doping \cite{dong}. It is shown that F doping shifts the
maximum of $\chi_0(q)$ from the M point of the Brillouin zone to an
IC {\bf q} vector, inducing thus an IC magnetic structure.
The appearance of an IC SDW upon
doping is also predicted in Ref.\cite{cvetko}
In view of the similarities between the 1111
and 122 families of pnictides, these results could also hold for the latter
family,
accounting for the fact that a very small doping level ($x$=0.014) is enough
to push the material towards an IC SDW.

\subsection{The hyperfine constant and the Fe moment}
The saturated hyperfine field in BaFe$_2$As$_2$ is 5.4\,T
\cite{rotter} and the maximum hyperfine field at 4.2\,K in the Co doped
compounds has comparable magnitude (see insert of Fig.\ref{mod4k2}). These are
unusually small values for Fe, even in an itinerant magnet.
\begin{table}
\caption{Saturated Fe moment $m$ and hyperfine field $H_{hf}$, and the
ratio $r=H_{\rm hf}/m$ in AFe$_2$As$_2$ for A= Ba, Ca and Sr.}
\label{tchf}
\begin{center}
\begin{tabular}{||c|c|c|c||} \hline
  A                 & Ba   & Ca   & Sr  \\ \hline
$m$($\mu_B$)  & 0.87\cite{huang} & 0.80\cite{gold} & 1.0\cite{kan} \\ \hline
$H_{\rm hf}$(T)& 5.4\cite{rotter0} &10\cite{kumar} & 8.9\cite{tegel} \\ \hline
$r$(T/$\mu_B$) & 6.3  & 12.5 & 8.9 \\ \hline
\end{tabular}
\end{center}
\end{table}
In intermetallic
compounds, it is usually assumed that the Fermi contact interaction with
$s$-electrons is the dominant hyperfine coupling, implying that the Fe moment
{\bf m} is proportional to the hyperfine field {\bf H$_{hf}$}
\cite{freewat}. This is true in $\alpha$-Fe, with a hyperfine constant
$C_{\rm hf}$=$H_{\rm hf}/m \simeq$ 15\,T/$\mu_B$, and in a number of other
intermetallics with similar $C_{\rm hf}$ value. In the 122
family of isostructural pnictides, the ratios of hyperfine field and moment
are smaller and present rather scattered values (see Table
\ref{tchf}, where the moment is taken from neutron diffraction data).
A strict proportionality between moment and hyperfine field can be however
questioned in the pnictides. Indeed, the magnetic susceptibility has been
found to be anisotropic ($\chi_\perp/\chi_{//} \simeq$ 1.5-2) both
in undoped BaFe$_2$As$_2$ \cite{wang} and in Co-doped compounds
\cite{fra}. This is a hint to the
presence of an ``unquenched'' orbital moment at the Fe site. This implies a
probably small, but non-zero contribution of the $d$-orbitals to
the hyperfine field, with a sign opposite to that of the Fermi contact field.
Then, like in Fe$^{2+}$ compounds \cite{imbert}, the hyperfine field is no
longer strictly proportional to the moment. The hyperfine field modulation
then would not reflect exactly that of the Fe moment.

However, assuming as a first good approximation that proportionality holds in
the Ba(Fe$_{1-x}$Co$_x$)$_2$As$_2$
series with the same constant as in BaFe$_2$As$_2$, then the maximum Fe moment
of the modulation at 4.2\,K would be 0.96, 0.84 and 0.72\,$\mu_B$ for
$x$=0.014, 0.03 and $x$=0.045 respectively.

\subsection{Thermal variation of the moment and nature of the SDW transition}
\begin{figure}
\includegraphics[width=\linewidth]{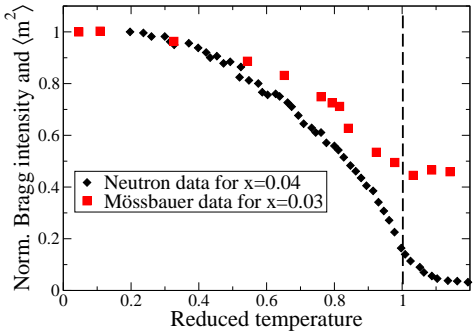}
\caption{[color online] Variation of the normalised values of the Bragg peak
intensity for $x$=0.04 from Ref.\cite{christianson} (diamonds) and of the
M\"ossbauer derived $\langle m^2(x,T) \rangle$ for $x$=0.03 from
Eqn.\ref{mneutr} (squares), as a function of the reduced temperature
$T/T_{\rm SDW}$. The M\"ossbauer data above $T_{\rm SDW}$ belong to the
short range AF regions.}
\label{intnorm}
\end{figure}

In the neutron diffraction studies of the series
Ba(Fe$_{1-x}$Co$_x$)$_2$As$_2$ \cite{huang,pratt,christianson,lester},
the integrated intensity of the magnetic Bragg peaks continuously drops
towards zero as temperature is increased towards $T_{\rm SDW}$.
This behaviour is in contrast with the $^{75}$As NMR work in BaFe$_2$As$_2$
\cite{kita}, where the transferred hyperfine field at the As nucleus shows a
weak thermal variation and it was concluded that the SDW transition is first
order. A similar discrepancy is observed in the present work in Co-doped
materials, where the main Fourier component $h_1$ of the modulation retains a
sizeable value at $T_{\rm SDW}$ (see Fig.\ref{pourc3} Top). However, a correct
comparison with the neutron data must take into
account the fact that the intensity of a magnetic Bragg peak is proportional
to the square of the magnetic moment. In case of a modulated structure with
Fourier coefficients $m_{2k+1}$, it should scale with $\langle m^2 \rangle$,
where the brackets denote an average over a period of the moment modulation:
\begin{equation}
\langle m^2(x,T) \rangle = \frac{1}{2} \sum_{k=0,n} m_{2k+1}^2(x,T) \propto
 \frac{1}{2} \sum_{k=0,n} h_{2k+1}^2(x,T).
\label{mneutr}
\end{equation}
Figure \ref{intnorm} shows the thermal variations of
$\langle m^2(x,T) \rangle$ (from the $h_{2k+1}(x,T)$ values
determined by M\"ossbauer spectroscopy) for $x$=0.03 and of the intensity of
a magnetic Bragg peak for $x$=0.04 \cite{christianson}. The $\langle m^2
\rangle$ value is reduced by half at $T_{\rm SDW}$, in disagreement with the
neutron data. Similar behaviours occur for $x$=0.014 and 0.045. It can thus be
concluded that the SDW transition in the Ba(Fe$_{1-x}$Co$_x$)$_2$As$_2$
series, as observed with the M\"ossbauer local probe, also presents a strong
first order character.

Comparing the moment values obtained from
the neutron data \cite{lester} for $x$=0.025 and 0.045 with the M\"ossbauer
results is more difficult since, in the scaling of the neutron Bragg
intensity, it was assumed that the magnetic structure is identical to that
of BaFe$_2$As$_2$, i.e. commensurate with the lattice spacings.
The mean saturated Fe moment value $m_s$ obtained from neutron diffraction
should be compared with:
\begin{equation}
m_s(x) = \sqrt{\langle m^2(x,T=4.2\,K) \rangle}.
\label{ms}
\end{equation}
Assuming proportionality between hyperfine field and moment, and using the
M\"ossbauer derived values for $h_{2k+1}(x,T=4.2\,K)$, one gets: $m_s$=0.70,
0.59 and 0.45\,$\mu_B$ resp. for $x$=0.014, 0.03 and 0.045. This must be
compared with $m_s$=0.35 and 0.17\,$\mu_B$ resp. for $x$=0.025 and 0.045
derived from the neutron data in Ref.\cite{lester}.

Further work is needed to resolve the disagreement between
local probe and neutron diffraction data about the Fe moment variation
with temperature and Co doping.

\section{Conclusions}

$^{57}$Fe M\"ossbauer absorption spectra have been measured in the
Ba(Fe$_{1-x}$Co$_x$)$_2$As$_2$ series for $x$=0.014, 0.03 and 0.045. In all
the samples, unusual magnetic hyperfine lineshapes are observed from 4.2\,K
to $T_{\rm SDW}$. They are interpreted in terms of a continuous
modulation of hyperfine fields at the Fe sites arising from an incommensurate
electronic SDW state. Incommensurability is present at the
lowest doping level, i.e. for $x$=0.014, while the magnetic structure in
BaFe$_2$As$_2$ is commensurate. Above $T_{\rm SDW}$, we do not observe a
purely paramagnetic signal, but a coexistence of magnetic
hyperfine and paramagnetic subspectra in a large temperature range, up
to  $T_{\rm AF} \sim 1.5 \ T_{\rm SDW}$ approximately for all samples. We
believe this is due to the persistence of AF short range correlated regions,
probably dynamic, well above the SDW transition, which is found to have a
strong first order character. Our
analysis of the hyperfine field modulations as a function of $x$ shows that
magnetism is not strongly depleted as doping increases, i.e. the mean Fe
moment or the maximum of the modulation does not decrease with $x$ as rapidly
as inferred from neutron diffraction data. We believe the
incommensurability of the magnetic structure
has not been evidenced by neutron scattering up to now due to the probable
very small value of the incommensurate component(s) of the SDW wave vector.
This makes M\"ossbauer spectroscopy a valuable tool to investigate these
structures: although the propagation vector cannot be obtained with this
technique, the spectral shape is sensitive to minute incommensurabilities.

\acknowledgments
We thank J. Bobroff and H. Alloul for fruitful discussions.



\begin{thebibliography}{24}
\expandafter\ifx\csname natexlab\endcsname\relax\def\natexlab#1{#1}\fi
\expandafter\ifx\csname bibnamefont\endcsname\relax
  \def\bibnamefont#1{#1}\fi
\expandafter\ifx\csname bibfnamefont\endcsname\relax
  \def\bibfnamefont#1{#1}\fi
\expandafter\ifx\csname citenamefont\endcsname\relax
  \def\citenamefont#1{#1}\fi
\expandafter\ifx\csname url\endcsname\relax
  \def\url#1{\texttt{#1}}\fi
\expandafter\ifx\csname urlprefix\endcsname\relax\def\urlprefix{URL }\fi
\providecommand{\bibinfo}[2]{#2}
\providecommand{\eprint}[2][]{\url{#2}}

\bibitem{kami}
\bibinfo{author}{\bibnamefont{Kamihara}~\bibfnamefont{Y. \emph{et~al}.}},
 \bibinfo{journal}{J. Am. Chem. Soc.} \textbf{\bibinfo{volume}{130}},
  \bibinfo{pages}{3296} (\bibinfo{year}{2008}).


\bibitem{sefat}
\bibinfo{author}{\bibfnamefont{Sefat}~\bibnamefont{A.S. \emph{et~al}.}},
 \bibinfo{journal}{Phys. Rev. Lett.} \textbf{\bibinfo{volume}{101}},
  \bibinfo{pages}{117004} (\bibinfo{year}{2008}).


\bibitem{chu}
\bibinfo{author}{\bibfnamefont{Chu}~\bibnamefont{J.H. \emph{et~al}.}},
 \bibinfo{journal}{Phys. Rev. B} \textbf{\bibinfo{volume}{79}},
  \bibinfo{pages}{014506} (\bibinfo{year}{2009}).


\bibitem{flo}
\bibinfo{author}{\bibfnamefont{Rullier-Albenque}~\bibnamefont{F.}},
\bibinfo{author}{\bibfnamefont{Colson}~\bibnamefont{D.}},
\bibinfo{author}{\bibfnamefont{Forget}~\bibnamefont{A.}},
\bibinfo{author}{\bibfnamefont{Alloul}~\bibnamefont{H.}},
 \bibinfo{journal}{Phys. Rev. Lett.} \textbf{\bibinfo{volume}{103}},
  \bibinfo{pages}{057001} (\bibinfo{year}{2009}).


\bibitem{huang}
\bibinfo{author}{\bibfnamefont{Huang}~\bibnamefont{Q. \emph{et~al}.}},
 \bibinfo{journal}{Phys. Rev. Lett.} \textbf{\bibinfo{volume}{101}},
  \bibinfo{pages}{257003} (\bibinfo{year}{2008}).


\bibitem{rotter0}
\bibinfo{author}{\bibfnamefont{Rotter}~\bibnamefont{M. \emph{et~al}.}},
 \bibinfo{journal}{Phys. Rev. B} \textbf{\bibinfo{volume}{78}},
  \bibinfo{pages}{020503} (\bibinfo{year}{2008}).


\bibitem{rotter}
\bibinfo{author}{\bibfnamefont{Rotter}~\bibnamefont{M. \emph{et~al}.}},
 \bibinfo{journal}{New Journal of Physics} \textbf{\bibinfo{volume}{11}},
  \bibinfo{pages}{025014} (\bibinfo{year}{2009}).

\bibitem{kita}
\bibinfo{author}{\bibfnamefont{Kitagawa}~\bibnamefont{K. \emph{et~al}.}},
 \bibinfo{journal}{J. Phys. Soc. Jpn.} \textbf{\bibinfo{volume}{77}},
  \bibinfo{pages}{114709} (\bibinfo{year}{2008}).

\bibitem{pratt}
\bibinfo{author}{\bibfnamefont{Pratt}~\bibnamefont{D. K. \emph{et~al}.}},
 \bibinfo{journal}{Phys. Rev. Lett.} \textbf{\bibinfo{volume}{103}},
  \bibinfo{pages}{087001} (\bibinfo{year}{2009}).

\bibitem{christianson}
\bibinfo{author}{\bibfnamefont{Christianson}~\bibnamefont{A. D. \emph{et~al}.}},
 \bibinfo{journal}{Phys. Rev. Lett.} \textbf{\bibinfo{volume}{103}},
  \bibinfo{pages}{087002} (\bibinfo{year}{2009}).

\bibitem{lester}
\bibinfo{author}{\bibfnamefont{Lester}~\bibnamefont{C. \emph{et~al}.}},
 \bibinfo{journal}{Phys. Rev. B} \textbf{\bibinfo{volume}{79}},
  \bibinfo{pages}{144523} (\bibinfo{year}{2009}).

\bibitem{lapl}
\bibinfo{author}{\bibfnamefont{Laplace}~\bibnamefont{Y. \emph{et~al}.}},
 \bibinfo{journal}{Phys. Rev. B} \textbf{\bibinfo{volume}{80}},
  \bibinfo{pages}{140501(R)} (\bibinfo{year}{2009}).

\bibitem{lapl2}
\bibinfo{author}{\bibfnamefont{Laplace}~\bibnamefont{Y. \emph{et~al}.}},
 \bibinfo{journal}{arXiv:0907.3973} (\bibinfo{year}{2009}).

\bibitem{ning}
\bibinfo{author}{\bibfnamefont{Ning}~\bibnamefont{F. L. \emph{et~al}.}},
 \bibinfo{journal}{Phys. Rev. B} \textbf{\bibinfo{volume}{79}},
  \bibinfo{pages}{140506(R)} (\bibinfo{year}{2009}).

\bibitem{wang}
\bibinfo{author}{\bibfnamefont{Wang}~\bibnamefont{X. F. \emph{et~al}.}},
 \bibinfo{journal}{Phys. Rev. Lett.} \textbf{\bibinfo{volume}{102}},
  \bibinfo{pages}{117005} (\bibinfo{year}{2009}).

\bibitem{zhang}
\bibinfo{author}{\bibfnamefont{Zhang}~\bibnamefont{G. M. \emph{et~al}.}},
 \bibinfo{journal}{EPL} \textbf{\bibinfo{volume}{86}},
  \bibinfo{pages}{37006} (\bibinfo{year}{2009}).

\bibitem{gold}
\bibinfo{author}{\bibfnamefont{Goldman}~\bibnamefont{A. I. \emph{et~al}.}},
 \bibinfo{journal}{Phys. Rev. B} \textbf{\bibinfo{volume}{78}},
  \bibinfo{pages}{100506} (\bibinfo{year}{2008}).

\bibitem{kan}
\bibinfo{author}{\bibfnamefont{Kaneko}~\bibnamefont{K. \emph{et~al}.}},
 \bibinfo{journal}{Phys. Rev. B} \textbf{\bibinfo{volume}{78}},
  \bibinfo{pages}{212502} (\bibinfo{year}{2008}).

\bibitem{kumar}
\bibinfo{author}{\bibfnamefont{Kumar}~\bibnamefont{K. \emph{et~al}.}},
 \bibinfo{journal}{Phys. Rev. B} \textbf{\bibinfo{volume}{79}},
  \bibinfo{pages}{012504} (\bibinfo{year}{2009}).


\bibitem{tegel}
\bibinfo{author}{\bibfnamefont{Tegel}~\bibnamefont{M. \emph{et~al}.}},
 \bibinfo{journal}{J. Phys.: Condens. Matter} \textbf{\bibinfo{volume}{20}},
  \bibinfo{pages}{452201} (\bibinfo{year}{2008}).

\bibitem{dong}
\bibinfo{author}{\bibfnamefont{Dong}~\bibnamefont{J. \emph{et~al}.}},
 \bibinfo{journal}{EPL} \textbf{\bibinfo{volume}{83}},
  \bibinfo{pages}{27006} (\bibinfo{year}{2008}).


\bibitem{freewat}
\bibinfo{author}{\bibfnamefont{Freeman}~\bibnamefont{A. J.}} \bibnamefont{and}
  \bibinfo{author}{\bibfnamefont{Watson}~\bibnamefont{R. E.}},
  \emph{\bibinfo{book}{Magnetism}}, \bibinfo{editor}{edited by G. T. Rado and H. Suhl}, Vol. \textbf{\bibinfo{volume}{IIA}} (Academic, New York),
  \bibinfo{year}{1965}, p.\bibinfo{pages}{167}.


\bibitem{fra}
\bibinfo{author}{\bibfnamefont{Rullier-Albenque}~\bibnamefont{F., \emph{private communication}}} (\bibinfo{year}{2009})


\bibitem{imbert}
\bibinfo{author}{\bibfnamefont{Hartmann-Boutron}~\bibnamefont{F.}} \bibnamefont{and}
  \bibinfo{author}{\bibfnamefont{Imbert}~\bibnamefont{P.}},
 \bibinfo{journal}{J. Appl. Phys.} \textbf{\bibinfo{volume}{39}},
  \bibinfo{pages}{775} (\bibinfo{year}{1968}).



\end{thebibliography}
\end{document}